\documentclass[11pt]{article}
\usepackage[colorlinks=true, allcolors=blue,linkcolor=blue, citecolor=blue]{hyperref}
\usepackage{sigsam, amsmath}
\usepackage{amsthm}
\usepackage{amssymb}
\usepackage{amsfonts}
\usepackage{mathtools}
\usepackage{maple}
\usepackage{orcidlink}
\usepackage[utf8]{inputenc}
\usepackage{breqn}
\usepackage{listings}
\usepackage{moreverb}
\usepackage{verbatim}
\usepackage{sverb}
\usepackage{xcolor}
\usepackage{url}
\usepackage{algorithm}
\usepackage{algpseudocode}
\usepackage{algorithmicx}
\usepackage{cleveref}
\newtheorem{theorem}{Theorem}
\newtheorem{definition}[theorem]{Definition}
\newtheorem{example}[subsection]{Example}
\newtheorem{remark}[subsection]{Remark}
\newtheorem{proposition}[theorem]{Proposition}

\providecommand{\keywords}[1]{\textbf{\textit{Keywords:}} #1}

\newcommand*{\QEDA}{\hfill\ensuremath{\blacksquare}}

\newcommand{\mmod}[1]{\ \mathrm{mod}\ #1}
\newcommand{\mfoldind}[3][n\equiv]{\chi_{\{#1 #2\mmod #3\}}}

\newcommand{\NN}{\mathbb{N}}

\newcommand{\QQ}{\mathbb{Q}}
\newcommand{\KK}{\mathbb{K}}

\DeclareMathOperator{\lcm}{lcm}

\issue{ISSAC'24}
\articlehead{Software paper} 
\titlehead{Computing with Hypergeometric-Type Terms}
\authorhead{B. Teguia Tabuguia}
\begin{document}
	
	\title{Computing with Hypergeometric-Type Terms}
	
	\author{Bertrand Teguia Tabuguia\\
		Department of Computer Science, University of Oxford, UK\\
		{\tt bertrand.teguia@cs.ox.ac.uk}}
	
	\date{}
	
	\maketitle
	
	\begin{abstract}
		\noindent Take a multiplicative monoid of sequences in which the multiplication is given by Hadamard product.  The set of linear combinations of interleaving monoid elements then yields a ring.  For hypergeometric sequences, the resulting ring is a subring of the ring of holonomic sequences. We present two algorithms in this setting: one for computing holonomic recurrence equations from hypergeometric-type normal forms and the other for finding products of hypergeometric-type terms. These are newly implemented commands in our Maple package \texttt{HyperTypeSeq}, available at \url{https://github.com/T3gu1a/HyperTypeSeq}, which we also describe.
	\end{abstract}
	
	
\section{Introduction}\label{sec:intro}
In \cite{abramovmsparse}, Abramov defined $m$-sparse sequences as those whose frequent non-zero terms have the same index modulo some positive integer $m$. Any such sequence has a general term of the form

\vspace{-0.25cm}
\begin{equation}\label{eq:msparse}
    c_n\, \mfoldind{j}{m},
\end{equation}
\vspace{-0.25cm}

\noindent where $(c)_n\coloneqq (c_n)_{n\in\NN}=(c_0,c_1,\ldots)$ is some non-zero sequence, and $\mfoldind{j}{m}$ evaluates the indicator function of the set $\{j+m\,k,\, k\in\NN\}$ at $n$, $j\in\NN,j<m$. We introduced the latter in \cite{teguia2023hypergeometric} under the name $m$-fold indicator sequence to enable a symbolic treatment of interlacement (also known as interleaving). See \cite[Definition 2]{teguia2023hypergeometric} for a formal definition. Note that we use the same notations from \cite{teguia2023hypergeometric}. Throughout this paper, $\KK$ denotes a field of characteristic zero. 

By convention, the zero sequence $(0,0,\ldots)$ is $(\mfoldind[]{0}{0})_n$ and the one sequence $(1,1,\ldots)$ is $(\mfoldind[]{0}{1})_n$. Although there is a similar understanding of interlacement in the literature (see \cite[Section 2.2]{KauersDfinite}), we are unaware of any algorithms or software to perform automatic computations when interlacements occur in the general term of a sequence. We consider linear combinations of interlaced sequences. These linear combinations define a ring when the corresponding sequences are closed under Hadamard product \cite[Theorem 1]{teguia2023hypergeometric}. We are interested in the case of hypergeometric sequences.
\begin{definition}[Hypergeometric-type sequence]\label{def:hyptypseq} A sequence $(s)_n$ is said to be of hypergeometric type if there exist finitely many $m$-fold indicator sequences $(\mfoldind[]{j_1}{m_1})_n,\ldots,(\mfoldind[]{j_l}{m_l})_n$ such that its general term $s_n$ writes
\begin{equation}\label{eq:hyptypseq}
    s_n = H_1(\sigma_1(n))\cdot \mfoldind{j_1}{m_1} + H_2(\sigma_2(n))\cdot \mfoldind{j_2}{m_2} + \cdots + H_l(\sigma_l(n))\cdot \mfoldind{j_l}{m_l},
\end{equation}
where $\sigma_i\colon \NN \longrightarrow \QQ$ is such that $\sigma_i(m_i\cdot n + j_i)\in\NN$, and $H_i(n)$ is a $\KK$-linear combination of hypergeometric terms, $i=1,\ldots,l$. We call the $H_i$'s the coefficients of $s_n$.
\end{definition}
\begin{example}[OEIS \href{https://oeis.org/A212579}{A212579} (see \cite{teguia2023hypergeometric})] The sequence $(s)_n$ whose general term $s_n$ counts the number of $(w,x,y,z)$ in $\{1,...,n\}^4$ such that $\min\left(|w-x|,|w-y|\right)=\min\left(|x-y|,|x-z|\right)$ is of hypergeometric type. We have
\begin{dmath}\label{maple1}
s_n = \frac{4}{9}+\frac{31}{12} n-3 n^{2}+\frac{67}{36} n^{3}-\frac{1}{4} n \mfoldind{0}{2}-\frac{4}{9} \mfoldind{0}{3}-\frac{8}{9} \mfoldind{1}{3}. 
\end{dmath} \QEDA
\end{example}
A hypergeometric-type term written in the form of \eqref{eq:hyptypseq} is in normal form. We denote by $(\mathcal{H}_T)$ the set of hypergeometric-type sequences, and $\mathcal{H}_T$ the one of their general terms in normal forms. We present our Maple software package \texttt{HyperTypeSeq} available on GitHub at \url{https://github.com/T3gu1a/HyperTypeSeq}. All computations of this paper can be done with Maple versions 2019-2023. The initial version of the package was made with the following commands.
\begin{itemize}
    \item $\texttt{mfoldInd}(\texttt{n},\texttt{m},\texttt{j})$: evaluates an $m$-fold indicator term or write it symbolically as $\chi_{\lbrace \mathit{modp} \left(n,m\right)=j \rbrace}.$
    \item $\texttt{HTSeval}(\texttt{s},\texttt{n}=\texttt{j})$: evaluates a hypergeometric-type term at index $\texttt{j}$.
    \item $\texttt{HTS}(\texttt{expr},\texttt{n})$: writes a given expression into a hypergeometric-type normal form whenever possible.
    \item $\texttt{REtoHTS}(\texttt{RE},\texttt{a}(\texttt{n}),\texttt{P})$: aims to decide whether a given holonomic term is of hypergeometric type or not by writing it in hypergeometric-type normal form. The argument $\texttt{P}$ is a procedure to compute the sequence terms from the recurrence or a list of sufficiently many initial terms.
    \item $\texttt{HolonomicRE}(\texttt{expr},\texttt{a}(\texttt{n}),\texttt{maxreorder}=\texttt{d})$: adapts $\texttt{HolonomicDE}$ from \cite{FPS} to search for a holonomic recurrence equation from an expression and a given order bound $\texttt{d}$. In \Cref{sec:holorec}, we present an extension of this implementation with \Cref{algo:Algo1} to enable hypergeometric-type terms as inputs.
\end{itemize}
\begin{example}\label{ex:maple1} Let us give some examples. The next section will give examples for \texttt{HolonomicRE}.
\begin{lstlisting}
> with(HyperTypeSeq): #load the package
> mfoldInd(7,3,1); mfoldInd(n,3,1);
s:=n!*mfoldInd(n,4,2)+2^n*mfoldInd(n,2,1); HTSeval(s,n=6);
\end{lstlisting}

\vspace{-0.8cm}

\begin{align}
&1\label{maple11}\\
&\chi_{\left\{\mathit{modp} \left(n,3\right)=1\right\}} \label{maple12}\\
&s\coloneqq n! \chi_{\left\{\mathit{modp} \left(n,4\right)=2\right\}}+2^{n} \chi_{\left\{\mathit{modp} \left(n,2\right)=1\right\}}\label{maple13}\\
&720 \label{maple14}
\end{align}

\vspace{-0.5cm}

\begin{lstlisting}
> HTS(cos(n*Pi/4)^2, n)
\end{lstlisting}

\vspace{-0.4cm}

\begin{dmath}\label{maple15}
\frac{1}{2}+\frac{\left(-1\right)^{\frac{n}{2}} \chi_{\left\{\mathit{modp} \left(n,2\right)=0\right\}}}{2}
\end{dmath}

\vspace{-0.25cm}

\begin{lstlisting}
> RE:= (n + 1)*(n^6 + 14*n^5 + 35*n^4 - 350*n^3 - 2436*n^2 - 5545*n - 4319)*(n - 2)*a(n) - (n - 1)*(n^6 + 8*n^5 - 20*n^4 - 370*n^3 - 1301*n^2 - 1799*n - 838)*a(n + 1) - (n + 1)*(n - 7)*(n^6 + 14*n^5 + 35*n^4 - 350*n^3 - 2436*n^2 - 5545*n - 4319)*a(n + 5) + (n^6 + 8*n^5 - 20*n^4 - 370*n^3 - 1301*n^2 - 1799*n - 838)*(n - 6)*a(n + 6) = 0:
> L := [1, 7, 2, 6, 24, 120, 721]:
> REtoHTS(RE, a(n), L)
\end{lstlisting}
\begin{dmath}\label{maple16}
n !-\left(n -7\right) \chi_{\left\{\mathit{modp} \left(n ,5\right)=1\right\}}
\end{dmath}\QEDA
\end{example}

The commands $\texttt{HTS}$ and $\texttt{REtoHTS}$ rely on the Maple command \texttt{LREtools:-mhypergeomsols} from \texttt{FPS} \cite{FPS}, which implements the $m$-fold hypergeometric term solver from \cite{BTWKsymbconv}. Of course, $\texttt{HTS}$ also uses \texttt{HolonomicRE} to convert expressions into P-recursive equations. 

\section{Recurrence Equations}\label{sec:holorec}

The proof of \cite[Proposition 3]{teguia2023hypergeometric} is constructive and can be turned into the following algorithm.

\begin{algorithm}[ht]\caption{P-recursive equations for terms in $\mathcal{H}_T$}\label{algo:Algo1}
    \begin{algorithmic} 
    \\ \Require $s_n\coloneqq \sum_{i=1}^l H_i(\sigma_i(n))\,\mfoldind{j_i}{m_i} \in \mathcal{H}_T$.
    \Ensure A $P$-recursive equation satisfied by $s_n$.
    \begin{enumerate}
        \item Set $u_i(m_i\,n+j_i)\coloneqq H_i(\sigma_i(m_i\,n+j_i)),$ $i=1,\ldots,l$.
        \item\label{alg:step2} Find a P-recursive $m_i$-shift recurrence equation satisfied by $u_i(m_i\,n+j_i)$, $i=1,\ldots,l$.
        \item Deduce a P-recursive equation satisfied by $u_i(n)$ by the substitution $n\rightarrow \frac{n-j_i}{m_i}$, $i=1,\ldots,l$.
        \item\label{alg:step4} Return the P-recursive equation obtained by applying the P-recursive addition algorithm to $\sum_{i=1}^l u_i$.
    \end{enumerate}	
    \end{algorithmic}
\end{algorithm}

\vspace{-0.5cm}

\begin{remark} \Cref{algo:Algo1} is implemented by \texttt{HolonomicRE} for expressions containing $m$-fold indicator terms. For step \ref{alg:step4}, we use \texttt{AddHolonomicRE} from the subpackage \texttt{AlgebraHolonomicSeq} of \texttt{HyperTypeSeq}. This procedure uses a linear dynamical system defined by the equations and proceeds like our standard \texttt{HolonomicRE} to find the desired equation. One can also recursively use the implementation in the \texttt{GFUN} package \cite{salvy1994gfun}.
\end{remark}
\begin{example}\label{ex:maple2}\item 
\begin{lstlisting}
> with(HyperTypeSeq):
> HolonomicRE(3^n*mfoldInd(n,3,1)+(2^n+n)*mfoldInd(n,2,0),a(n))
\end{lstlisting}

\vspace{-0.35cm}

\begin{dmath}\label{maple21}
\left(-353808 n^{2}-586764 n-150444\right) a\! \left(n\right)-78732\, a\! \left(n+1\right)+\left(442260 n^{2}+25839 n-44901\right) a\! \left(n+2\right)+\left(13104 n^{2}+21732 n+25255\right) a\! \left(n+3\right)+\left(-88452 n^{2}+30213 n+23544\right) a\! \left(n+4\right)+\left(-16380 n^{2}-957 n+1663\right) a\! \left(n+5\right)-729\, a\! \left(n+6\right)+\left(3276 n^{2}-1119 n-764\right) a\! \left(n+7\right)=0
\end{dmath}

\vspace{-0.3cm}

\begin{lstlisting}
> HolonomicRE(n!*mfoldInd(n, 4, 3) + pochhammer(2, n), a(n))
\end{lstlisting}

\vspace{-0.35cm}

\begin{dmath}\label{maple22}
\left(n+4\right) \left(n+3\right) \left(n+2\right) \left(n+1\right) \left(n+5\right) a\! \left(n\right)-\left(n+5\right) \left(n+4\right) \left(n+3\right) \left(n+2\right) a\! \left(n+1\right)+\left(-n-5\right) a\! \left(n+4\right)+a\! \left(n+5\right)=0 
\end{dmath}
\QEDA
\end{example}

\section{Products}\label{sec:prodhyp}

The product algorithm is essentially based on the following proposition, which can be seen as a consequence of the \textit{Chinese Remainder Theorem}. 

\begin{proposition}[see Lemma 2 in \cite{teguia2023hypergeometric}]\label{prop:prop1} Let $j_1,j_2,m_1,m_2\in\NN$ such that $j_1<m_1, j_2<m_2$. Let $\mu=\lcm(m_1,m_2)$ and 
\[\mathcal{N}= \left\{j<\mu, j=j_1+m_1\,k_1 = j_2+m_2\,k_2,\, k_1,k_2\in\NN\right\}.\]
Then $\mathcal{N}=\{j_0\}$ or $\mathcal{N}=\emptyset$ and we have
\begin{equation}
    \mfoldind{j_1}{m_1}\cdot \mfoldind{j_2}{m_2} = \begin{cases}
        \mfoldind{j_0}{\mu},\, \, \phantom{0=}\textnormal{if } \mathcal{N}=\{j_0\}\\[0.33mm]
        0=\mfoldind{0}{0},\, \, \textnormal{if } \mathcal{N}=\emptyset
    \end{cases}.
\end{equation}
\end{proposition}
On account of \Cref{prop:prop1}, one can easily implement an algorithm for computing products of hypergeometric-type terms. Our implementation is the command \texttt{HTSproduct} from \texttt{HyperTypeSeq}.

\begin{example}\label{ex:maple3} \item   
\begin{lstlisting}
> with(HyperTypeSeq):
> u1 := (n!)^2*mfoldInd(n, 3, 1) + n^3*mfoldInd(n, 2, 1); 
u2 := (n + 1)*mfoldInd(n, 4, 3)/n! + (n + 2)*mfoldInd(n, 2, 0)
\end{lstlisting}

\vspace{-0.7cm}

\begin{align}
&\mathit{u1} \coloneqq n !^{2} \chi_{\left\{\mathit{modp} \left(n ,3\right)=1\right\}}+n^{3} \chi_{\left\{\mathit{modp} \left(n ,2\right)=1\right\}} \label{maple31}\\
&\mathit{u2} \coloneqq \frac{\left(n +1\right) \chi_{\left\{\mathit{modp} \left(n ,4\right)=3\right\}}}{n !}+\left(n +2\right) \chi_{\left\{\mathit{modp} \left(n ,2\right)=0\right\}} \label{maple32}
\end{align}

\vspace{-0.4cm}

\begin{lstlisting}
> HTSproduct(u1, u2, n)
\end{lstlisting}

\vspace{-0.4cm}

\begin{dmath}\label{maple33}
\frac{n^{3} \left(n +1\right) \chi_{\left\{\mathit{modp} \left(n ,4\right)=3\right\}}}{n !}+n !^{2} \left(n +2\right) \chi_{\left\{\mathit{modp} \left(n ,6\right)=4\right\}}+n ! \left(n +1\right) \chi_{\left\{\mathit{modp} \left(n ,12\right)=7\right\}}
\end{dmath}
\QEDA
\end{example}

\section{Conclusion}

As a conclusion, we can state that the ring $(\mathcal{H}_T)$ is computable. We gave some arguments in \cite[Section 3.1]{teguia2023hypergeometric} on the reasons why canonical forms are difficult to define in $\mathcal{H}_T$. However, the algorithm implemented by \texttt{HTS} computes normal forms \cite[Chapter 3]{geddes1992algorithms}. This implies that we can recognize equivalent formulas. Indeed, for any two equivalent hypergeometric-type terms in $\mathcal{H}_T$, the normal form returned by \texttt{HTS} for their difference is zero. With the extension of \texttt{HolonomicRE} to hypergeometric-type input, and the implementation of the product algorithm \texttt{HTSproduct}, we can solve the zero-equivalence problem in $(\mathcal{H}_T)$.

\medskip

\section*{Acknowledgment} The author was supported by UKRI Frontier Research Grant EP/X033813/1.

\end{document}